\documentclass[aps, prd, onecolumn, showpacs, showkeys, preprint, preprintnumbers, nofootinbib, superscriptaddress]{revtex4-1}

\usepackage{amssymb,amsbsy,amsmath,amsfonts}
\usepackage{yhmath}

\usepackage{slashed}
\usepackage{bm}
\usepackage{times}

\usepackage[dvipsnames,table,xcdraw]{xcolor}
\definecolor{lightyellow}{RGB}{255,250,205}

\usepackage{graphicx}
\graphicspath{{./}{./img/}{./fig/}{./image/}{./figure/}{./picture/}}
\usepackage{epstopdf}

\usepackage{tabularx}
\usepackage{multirow}
\usepackage{longtable}

\usepackage{hyperref}
\hypersetup{
  colorlinks=true,        
  linkcolor=blue,         
  citecolor=cyan,         
  urlcolor=red,           
}

\begin{document}

\title{Possible bound states with hidden bottom from $\bar{K}^{(*)}B^{(*)}\bar{B}^{(*)}$ systems}

\author{Xiu-Lei Ren}
\email{xiulei.ren@rub.de}
\affiliation{Ruhr-Universit\"{a}t Bochum, Fakult\"{a}t f\"{u}r Physik und Astronomie, Institut f\"{u}r Theoretische Physik II, D-44780 Bochum, Germany}

\author{Zhi-Feng Sun}
\email{sunzf@lzu.edu.cn}
\affiliation{School of Physical Science and Technology \& Research Center for Hadron and CSR Physics,
Lanzhou University, Lanzhou 730000, China }

\begin{abstract}
 We study the three-body systems of $\bar{K}^{(*)}B^{(*)}\bar{B}^{(*)}$ by solving the Faddeev equations in the fixed-center approximation, where the light particle $\bar{K}^{(*)}$ interacts with the heavy bound states of $B\bar{B}$ ($B^*\bar{B}^*$) forming the clusters. In terms of the very attractive $\bar{K}^*B$ and $\bar{K}^*B^*$ subsystems, which are constrained by the observed $B_{s1}(5830)$ and $B_{s2}^*(5840)$ states in experiment, we find two deep bound states, containing the hidden-bottom components, with masses $11002\pm 63$ MeV and $11078\pm 57$ MeV in the $\bar{K}^*B\bar{B}$ and $\bar{K}^*B^*\bar{B}^*$ systems, respectively. The two corresponding states with higher masses of the above systems are also predicted. 
 In addition, using the constrained two-body amplitudes of $\bar{K}B^{(*)}$ and $\bar{K}\bar{B}^{(*)}$ via the hidden gauge symmetry in the heavy-quark sector, we also find two three-body $\bar{K}B\bar{B}$ and $\bar{K}B^{*}\bar{B}^*$ bound states.    
\end{abstract}


\maketitle

\date{\today}

\section{Introduction}

With the development of experiments, a large number of hadronic states have been reported~\cite{Tanabashi:2018oca}, which provides an ideal playground to deepen our understanding of the nonperturbative  quantum chromodynamics (QCD). The interpretation of hadronic states is one of the most important issues in hadronic physics (see Refs.~\cite{Olsen:2017bmm,Guo:2017jvc,Ali:2017jda,Chen:2016qju,Chen:2016spr,Liu:2013waa} for reviews), particularly for the exotic states which cannot be easily collected as $q\bar{q}$ or $qqq$ states, e.g. the so-called $XYZ$ states. Recently, the exotic hadrons with the open/hidden heavy quark components, such as $Z_c(3900)$~\cite{Ablikim:2013mio,Liu:2013dau}, $Z_b(10510)$ and $Z_b(10560)$~\cite{Belle:2011aa}, $P_c(4380)$ and $P_c(4450)$~\cite{Aaij:2015tga},  have been reported and attracted great attention from the experimental and theoretical physicists.  Most of the heavy flavor meson resonances can be interpreted as the tetraquarks~\cite{Wu:2017weo,Wu:2016gas,Esposito:2014rxa} and/or the meson-meson molecules~\cite{Gamermann:2006nm,Molina:2009ct,Olsen:2017bmm,Dias:2014pva,Xiao:2013yca,Sun:2011uh,Sun:2012sy,Ohkoda:2012hv}. Besides, several heavy flavor mesons have been predicted in the three-body systems, 
like $\rho D^{(*)} \bar{D}^{(*)}$~\cite{Bayar:2015oea,Durkaya:2015wra}, $\rho B^*\bar{B}^*$~\cite{Bayar:2015zba}, $KDD$~\cite{MartinezTorres:2018zbl}, $KD\bar{D}^*$~\cite{Ma:2017ery,Ren:2018pcd}, $\bar{K}B\bar{B}^*$~\cite{Ma:2017ery}, $\bar{K}B^*B^*$~\cite{Valderrama:2018knt},
$D^*D^*\bar{D}^{(*)}$~\cite{Valderrama:2018sap}, 
$BDD$ ($BD\bar{D}$)~\cite{Dias:2017miz}, and $D^{(*)}B^{(*)}\bar{B}^{(*)}$~\cite{Dias:2018iuy}.

The standard method to study three-body systems refers to the Faddeev equations~\cite{Faddeev:1960su}. Since it is very difficult to solve exactly, one usually introduces some reasonable approximations of the Faddeev equations, such as the use of separable potentials and energy-independent kernels, the widely-used Alt-Grassberger-Sandras approach~\cite{Alt:1967fx}. 
Recently, a different approach to solve the  Faddeev equations was proposed to study the three-hadron systems~\cite{MartinezTorres:2007sr,Khemchandani:2008rk,MartinezTorres:2008gy}, which relies on the on-shell two-body scattering amplitudes. In addition, another approximation of the Faddeev equations, which is the so-called fixed-center approximation (FCA), has been employed in the studies of $\bar{K}d$ interaction at low energies~\cite{Chand:1962ec,Barrett:1999cw,Deloff:1999gc,Kamalov:2000iy}. In Refs.~\cite{Toker:1981zh,Gal:2006cw}, it is shown that the FCA is a rather good approximation, especially for the system with one light particle and one heavy cluster formed by the other two particles. 
Nowadays, the FCA has been applied in many problems~
\cite{Roca:2010tf,YamagataSekihara:2010qk, Ren:2018pcd,
Xiao:2012dw,Xiao:2015mqa,MartinezTorres:2010ax,Roca:2011br,
Liang:2013yta,Bayar:2013bta,Durkaya:2015wra,
Bayar:2015oea,Bayar:2015zba,MartinezTorres:2016ytd,Zhang:2016bmy,
Debastiani:2017vhv,Dias:2017miz,Dias:2018iuy,Xie:2010ig,Xie:2011uw,
Bayar:2011qj,Bayar:2012rk,Bayar:2012hn,Sekihara:2016vyd,Bayar:2012dd} and is accepted as a reasonable tool in the study of bound systems: such as the systems with three mesons: $\phi K\bar{K}$~\cite{MartinezTorres:2010ax}, $\eta K\bar{K}$ and $\eta' K\bar{K}$~\cite{Liang:2013yta},  $\rho K\bar{K}$~\cite{Bayar:2013bta}, $\rho D\bar{D}$~\cite{Durkaya:2015wra},  $\rho D^* \bar{D}^*$~\cite{Bayar:2015oea}, $\rho B^* B^*$~\cite{Bayar:2015zba}, $\eta' K\bar{K}$~\cite{MartinezTorres:2016ytd}, $\pi\bar{K}K^*$~\cite{Zhang:2016bmy}, $DKK$ and $DK\bar{K}$~\cite{Debastiani:2017vhv}, $KD\bar{D}^*$~\cite{Ren:2018pcd}, $BDD$ and $BD\bar{D}$~\cite{Dias:2017miz}, and $D^{(*)} B^{(*)} B^{(*)}$~\cite{Dias:2018iuy}; the systems with  multimesons: multi-$\rho$~\cite{Roca:2010tf}, $K^*$-multi-$\rho$~\cite{YamagataSekihara:2010qk},  $D^*$-multi-$\rho$~\cite{Xiao:2012dw}, $K$-multi-$\rho$~\cite{Xiao:2015mqa}; the systems with two mesons and one baryon: $N\bar{K}K$~\cite{Xie:2010ig}, $\pi\rho\Delta$~\cite{Xie:2011uw}; the systems with one meson and two baryons: $\bar{K}NN$~\cite{Bayar:2011qj,Bayar:2012rk,Bayar:2012hn,Sekihara:2016vyd}, $DNN$~\cite{Bayar:2012dd}. 
Among them, several possible bound states with the nature of open/hidden charm and bottom have been predicted, such as the six-quark state $K^*(4307)$ with the strange and hidden charm structure~\cite{Ren:2018pcd}. 

In the present work, we extend the study of Ref.~\cite{Ren:2018pcd} to the bottom sector to investigate the possible bound states from the $\bar{K}^{(*)} B^{(*)}\bar{B}^{(*)}$ systems with isospin $I=1/2$ using the fixed-center approximation of the Faddeev equations. We take $B \bar{B}$, $B^* \bar{B}^*$ as clusters (denoted as $R$ in the following) and $\bar{K}^{(*)}$ as a third particle to scatter,  which satisfies the general criteria of the FCA to the Faddeev equations: that the mass of the third particle $P_3$ should be smaller than a stable cluster (such as a bound state) 
composed of the two other particles $P_1$ and $P_2$. In Ref.~\cite{Ozpineci:2013qza}, a single bound state of $B\bar{B}$ system with $J^P=0^+$ and the three-degenerate states of $B^*\bar{B}^*$ system with $J^{P}=0^+,\, 1^+,\, 2^+$ was found in isospin 0 using the coupled-channel chiral unitary approach. 
In principle, the spin state of $B^*\bar{B}^*$ cluster can be chosen as $J_{R}=0$, $1$, or $2$. 
Similar to Ref.~\cite{Bayar:2015zba}, we are interested in the largest total spin of a molecular state from the $\bar{K}^{(*)}B^*\bar{B}^*$ systems, therefore, the $J_R=2$ cluster of $B^*\bar{B}^*$ is preferred. The corresponding wave function is simple with the spins of $B^*$ and $\bar{B}^*$ aligned, which will relatively simplify the practical  calculation. 
Besides, for the two-body subsystem, such as $\bar{K}^* B^*$, which can also produce the three spin bound states with $0^+$, $1^+$, and $2^+$, the spin-aligned $2^+$ state is more bound, by around 60 MeV, than the other spin states. Thus, the largest spin state of $\bar{K}^{(*)}B^*\bar{B}^*$  
would produce a larger binding than the other total spin systems. 
Therefore, we will investigate the $J=0$ $\bar{K} B\bar{B}$,   $J=1$ $\bar{K}^*B\bar{B}$, $J=2$ $\bar{K}B^*\bar{B}^*$ and $J=3$ $\bar{K}^*B^*\bar{B}^*$ systems with isospin $I=1/2$ and study the possibility to produce the bound states.

In the following Sec.~\ref{sec2}, we will first present the details of the formalism employed in the FCA framework. The input of two-body amplitudes are also calculated in the chiral unitary approach. The predicted bound states from $\bar{K}^{(*)}B^{(*)}\bar{B}^{(*)}$ are shown in Sec.~\ref{sec3} with the corresponding discussion. Finally, a summary is given in Sec.~\ref{sec4}.

\section{Theoretical Framework}\label{sec2}

\subsection{Fixed-center approximation to Faddeev equations}

\begin{figure}[b]
\begin{center}
\includegraphics[width=0.9\textwidth]{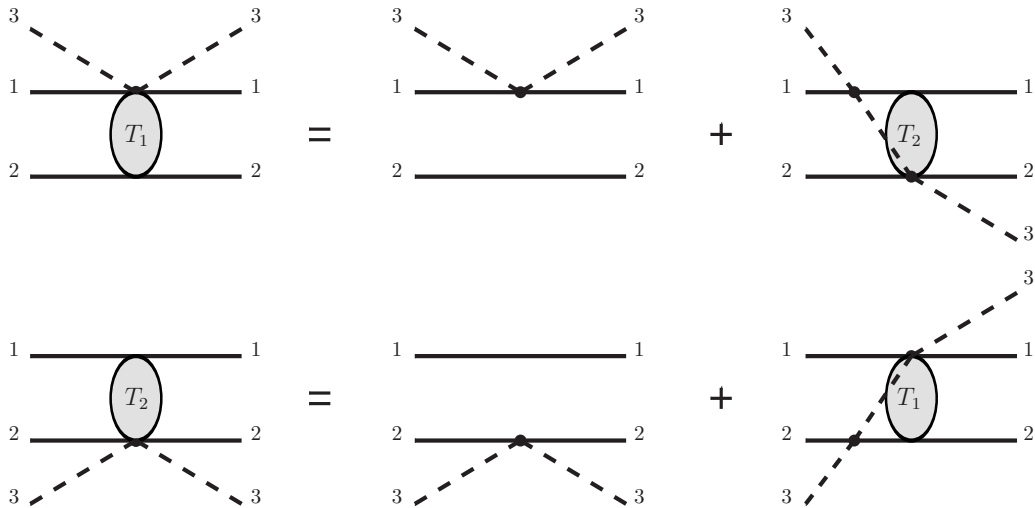}
\caption{Diagrammatic representation of the FCA to Faddeev equations.}
\label{Fig:FCA}
\end{center}
\end{figure}

Under the FCA, the total elastic scattering $T$-matrix of three-body systems can be simplified as the 
sum of two Faddeev partitions,  
\begin{equation}
  T_{\bar{K}^{(*)}R}  = T_1 + T_2\, ,
\end{equation}
where $T_1$ and $T_2$ describe the iterated interactions of the $\bar{K}^{(*)}$ scattering off the cluster $R$ with a first collision on $B$ ($B^*$) and $\bar{B}$ ($\bar{B}^*$), respectively. 
These interactions are illustrated in Fig.~\ref{Fig:FCA} and can be expressed as the two coupled equations, 
\begin{eqnarray}
  T_1 &=& t_{\bar{K}^{(*)}B^{(*)}} + t_{\bar{K}^{(*)}B^{(*)}}\, G_0 \,T_{2}\,,\\
  T_2 &=& t_{\bar{K}^{(*)}\bar{B}^{(*)}} + t_{\bar{K}^{(*)}\bar{B}^{(*)}} \, G_0 \, T_{1}\,,
 \end{eqnarray}
where the two-body scattering amplitudes $t_{\bar{K}^{(*)}B^{(*)}}$ and $t_{\bar{K}^{(*)}\bar{B}^{(*)}}$  denote the transition matrices for $\bar{K}^{(*)}B^{(*)}$ and  $\bar{K}^{(*)}\bar{B}^{(*)}$ elastic scattering in the isospin basis, respectively. The loop function $G_0$ is the Green function of the $\bar{K}^{(*)}$ meson propagating in the $B\bar{B}$ or $B^*\bar{B}^*$ cluster. 

In isospin space, the wave function of three-body system, combining the third particle $\bar{K}^{(*)}$ and the cluster $R$ ($B^{(*)}\bar{B}^{(*)}$ ), can be written as
\begin{equation}
  \left|\bar{K}^{(*)}R;\, I=\frac{1}{2},\, I_z=\frac{1}{2}\right\rangle = \left|I=\frac{1}{2}, I_z=\frac{1}{2}\right\rangle_{\bar{K}^{(*)}} 
  \otimes |I=0, I_z=0\rangle_{B^{(*)}\bar{B}^{(*)}}\,,
\end{equation} 
with 
\begin{equation}
 |I=0, I_z=0\rangle_{B^{(*)}\bar{B}^{(*)}} = \frac{1}{\sqrt{2}}\left(\left|\left(\frac{1}{2},-\frac{1}{2}\right)\right\rangle - \left|\left(-\frac{1}{2},\frac{1}{2}\right)\right\rangle\right),
\end{equation}
where the kets in the right-hand side indicate the $I_z$ components of the particles $B^*$ and $\bar{B}^*$ with $|(I_z^{B^*}, I_z^{\bar{B}^*})\rangle$. 
The corresponding amplitude $\langle \bar{K}^{(*)} R|\,t \,|\bar{K}^{(*)}R\rangle$ of single scattering in the isospin basis can be written as 
\begin{eqnarray}
  \langle \bar{K}^{(*)} R|\, t\, |\bar{K}^{(*)}R\rangle
   &=& \left( \left\langle I=\frac{1}{2},\, I_z=\frac{1}{2}\right|_{\bar{K}^{(*)}} \otimes  
   \left\langle I=0, I_z=0 \right|_{B^{(*)}\bar{B}^{(*)}} \right)  (t_{\bar{K}^{(*)}B^{(*)}} + t_{\bar{K}^{(*)}\bar{B}^{(*)}})   \nonumber\\
     &&  \left(\left|I=\frac{1}{2}, I_z=\frac{1}{2}\right\rangle_{\bar{K}^{(*)}} 
  \otimes |I=0, I_z=0\rangle_{B^{(*)}\bar{B}^{(*)}}\right)\nonumber\\
        &=& \left[\left\langle\frac{1}{2}\right| \otimes \frac{1}{\sqrt{2}} \left(\left\langle\left(\frac{1}{2},-\frac{1}{2}\right)\right| - \left\langle\left(-\frac{1}{2},\frac{1}{2}\right)\right| \right)\right] (t_{\bar{K}^{(*)}B^{(*)}} + t_{\bar{K}^{(*)}\bar{B}^{(*)}}) \nonumber\\
      &&\left[\left|\frac{1}{2}\right\rangle \otimes \frac{1}{\sqrt{2}} \left(\left|\left(\frac{1}{2},-\frac{1}{2}\right)\right\rangle - \left|\left(-\frac{1}{2},\frac{1}{2}\right)\right\rangle \right)\right]  \nonumber\\
        &=& \frac{1}{4}\left( 3\, t_{\bar{K}^{(*)}B^{(*)}}^{I=1} + t_{\bar{K}^{(*)}B^{(*)}}^{I=0} 
        +  3\,t_{\bar{K}^{(*)}\bar{B}^{(*)}}^{I=1} + t_{\bar{K}^{(*)}\bar{B}^{(*)}}^{I=0} \right).
\end{eqnarray}
One introduces two scattering $T$-matrices of the left/right collision, 
\begin{eqnarray}\label{Eq:twobodyamp}
 t_{\bar{K}^{(*)}B^{(*)}}  = \frac{1}{4}\left( 3\, t_{\bar{K}^{(*)}B^{(*)}}^{I=1} + t_{\bar{K}^{(*)}B^{(*)}}^{I=0} \right), && \quad 
 t_{\bar{K}^{(*)}\bar{B}^{(*)} } = \frac{1}{4}\left( 3\, t_{\bar{K}^{(*)}B^{(*)}}^{I=1} +  t_{\bar{K}^{(*)}B^{(*)}}^{I=0}\right),
 \end{eqnarray}
where the two-body amplitudes, $t_{\bar{K}^{(*)}B^{(*)}}^{I=0,1}$ and  $t_{\bar{K}^{(*)}\bar{B}^{(*)}}^{I=0,1}$, are the input of the FCA equations.

\subsection{Two-body amplitudes of subsystems}
In order to obtain the two-body amplitudes in Eq.~(\ref{Eq:twobodyamp}), we follow the calculation details in Ref.~\cite{Sun:2018zqs} and employ the lowest order Lagragians with the local hidden gauge symmetry in the SU(4) sector,~
\footnote{Note that the extension of the local hidden gauge approach from the light-quark sector~\cite{Ecker:1989yg,Nagahiro:2008cv} to the heavy-quark sector is possible if the heavy quarks of hadrons are just spectators and the major contributions of the interaction is from the exchange of light vector mesons ($\rho,~\omega,~\phi$). In this case, one can employ the SU(4) symmetry formally in the Lagrangians (e.g. Eq.~[\ref{Eq:lag})] and the actual SU(3) subgroup is used in the evaluation of the vertices. For more discussions and the proof, one can refer to Sec.~II of Ref.~\cite{Sakai:2017avl} and Sec. II and the Appendix of Ref.~\cite{Liang:2017ejq}.}  
\begin{eqnarray}\label{Eq:lag}
\mathcal{L}_\mathrm{eff.} &=& \mathcal{L}_\mathrm{PPPP} + \mathcal{L}_\mathrm{VVVV} + \mathcal{L}_\mathrm{VPP} + \mathcal{L}_\mathrm{VVV} \, \nonumber\\
   &=& -\frac{1}{24\,f_\pi^2} \langle [P,\,\partial_\mu P] [P,\,\partial^\mu P] \rangle + \frac{g^2}{2} \langle V_\mu V_\nu V^\mu V^\nu - V_\nu V_\mu V^\mu V^\nu \rangle \nonumber\\
   &&\quad - i\,g\,\langle V_\mu[P,\, \partial^\mu P]\rangle + i\,g\, \langle V_\mu[V_\nu,\, \partial^\mu V^\nu]\rangle,
\end{eqnarray}
where $f_\pi$ denotes the pion decay constant $f_\pi=93$ MeV and the coupling $g$ is determined through the SU(4) symmetry~\cite{Sun:2018zqs}. The fields of pseudoscalar mesons ($P$) and vector mesons ($V$)  are collected in the $4\times 4$ matrices, 
\begin{eqnarray}\label{matrix}
P&=&\left(\begin{array}{cccc}
\frac{\eta}{\sqrt{3}}+\frac{\eta^\prime}{\sqrt{6}}+\frac{\pi^0}{\sqrt{2}}&\pi^+&K^{+}&B^{+}\\
\pi^-&\frac{\eta}{\sqrt{3}}+\frac{\eta^\prime}{\sqrt{6}}-\frac{\pi^0}{\sqrt{2}}&K^{0}&B^{0}\\
K^{-}&\bar{K}^{0}&-\frac{\eta}{\sqrt{3}}+\sqrt{\frac{2}{3}}\eta^\prime &B_s^{0}\\
B^{-}&\bar{B}^{0}&\bar{B}_s^{0}&\eta_b
\end{array}\right)\,,\nonumber\\
V_\mu&=&\left(\begin{array}{cccc}
\frac{\omega}{\sqrt{2}}+\frac{\rho^0}{\sqrt{2}}&\rho^+&K^{*+}&B^{*+}\\
\rho^-&\frac{\omega}{\sqrt{2}}-\frac{\rho^0}{\sqrt{2}}&K^{*0}&B^{*0}\\
K^{*-}&\bar{K}^{*0}&\phi &B_s^{*0}\\
B^{*-}&\bar{B}^{*0}&\bar{B}_s^{*0}&\Upsilon
\end{array}\right)_\mu.
\end{eqnarray}

After considering the contact interactions and one-boson exchange contributions within the coupled-channel approach,
 the $s$-wave potentials,
$v_{\bar{K}^{(*)}B^{(*)}}$ and $v_{\bar{K}^{(*)}\bar{B}^{(*)}}$, are projected, as shown in Ref.~\cite{Sun:2018zqs}. 
To keep the self-consistency of the current work, we summarize all the needed two-body potentials in the Appendix.

In the chiral unitary approach, the interaction kernels $v_{\bar{K}^{(*)}B^{(*)}}$ or $v_{\bar{K}^{(*)}\bar{B}^{(*)}}$ can be resumed in the Bethe-Salpeter equation, 
\begin{equation}
   t = v + v\, G\, t = (1-v\,G)^{-1}\,v,
\end{equation} 
where $G$ is a diagonal matrix with the element being a two-meson loop function for the $i$th particle channel, 
\begin{equation}
  G(s_i) \, = \, i\, \int \frac{d^4 k}{(2\pi)^4} \frac{1}{k^2 - m_1^2 + i\epsilon} \frac{1}{(p-k)^2-m_2^2 + i\epsilon}\,,
\end{equation}
with the total four-momentum of two-meson system $p^\mu=(\sqrt{s_i},\, \bm{0})$ and $\sqrt{s_i}$ the center-of-mass (c.m.) energy. 
Using the cutoff regularization, the loop function changes as
\begin{equation}
  G(s_i) \, = \, \int_0^{k_\mathrm{max}} \frac{k^2\, d k}{(2\pi)^2} \frac{\omega_1 + \omega_2}{\omega_1\,\omega_2 \,[s_i^2 - (\omega_1+\omega_2)^2 + i\epsilon]}\,,
\end{equation}
where $k_\mathrm{max}$ denotes as the momentum cutoff and $\omega_i=\sqrt{k^2 + m_i^2}$.  
To evaluate the amplitudes $t_{\bar{K}B^{(*)}}^{I=0,1}$ and $t_{\bar{K}^*\bar{B}^{(*)}}^{I=0,1}$ , the momentum cutoff is chosen to be $k_\mathrm{max} = 1070$ MeV as given in Ref.~\cite{Sun:2018zqs}. 

 As demonstrated in Ref.~\cite{Roca:2010tf},  in the above calculation of two-body scattering, we use the normalization of Mandl and Shaw~\cite{Mandlbook}, which introduces different weight factors for the particle fields. Therefore, one has to consider these factors
in our two-body amplitudes 
\begin{eqnarray}
  \tilde{t}_{\bar{K}^{(*)}B^{(*)}} = \frac{2m_{R}}{2m_{B^{(*)}}} \, t_{\bar{K}^{(*)}B^{(*)}} , &\quad & \tilde{t}_{\bar{K}^{(*)}\bar{B}^{(*)}} = \frac{2m_{R}}{2m_{\bar{B}^{(*)}}} \, t_{\bar{K}^{(*)}\bar{B}^{(*)}}\,.
\end{eqnarray} 
Here we have taken the approximations $\frac{1}{\sqrt{2\omega_{B^{(*)}}}}=\frac{1}{\sqrt{2m_{B^{(*)}}}}$ and $\frac{1}{\sqrt{2\omega_{\bar{B}^{(*)}}}}=\frac{1}{\sqrt{2m_{\bar{B}^{(*)}}}}$, which is suitable for heavy-bottom particles as demonstrated in Ref.~\cite{Bayar:2015zba}.

\subsection{Total amplitude of three-body system}
Finally, we obtain the total amplitude $T_{\bar{K}^{(*)}R}$ of $\bar{K}^{(*)}B^{(*)}\bar{B}^{(*)}$ by solving the FCA Eqs.~(1-3), 
\begin{equation}\label{Eq:FCAamp}
  T_{\bar{K}^{(*)}R} = \frac{\tilde{t}_{\bar{K}^{(*)}B^{(*)}} + \tilde{t}_{\bar{K}^{(*)}\bar{B}^{(*)}} + 2\, \tilde{t}_{\bar{K}^{(*)}B^{(*)}}\,
   \tilde{t}_{\bar{K}^{(*)}\bar{B}^{(*)}}\, G_0 }
  {1-\tilde{t}_{\bar{K}^{(*)}B^{(*)}} \, \tilde{t}_{\bar{K}^{(*)}\bar{B}^{(*)}} \, G_0^2} ,
\end{equation}
which is apparently a function of the total invariant mass of the three-body system. The arguments in the two-body amplitudes, $t_{\bar{K}^{(*)}B^{(*)}}$ and $t_{\bar{K}^{(*)}\bar{B}^{(*)}}$, are  $s_1$ and $s_2$, which are (commonly) determined through
\begin{eqnarray}\label{Eq:MA}
  s_{1} &=& m_{\bar{K}^{(*)}}^2 + m_{B^{(*)}}^2 + \frac{m_{R}^2+m_{B^{(*)}}^2-m_{\bar{B}^{(*)}}^2}{2{m_{R}^2}}\left(s - m_{\bar{K}^{(*)}}^2 - m_{R}^2\right),\nonumber\\
  s_{2} &=& m_{\bar{K}^{(*)}}^2 + m_{\bar{B}^{(*)}}^2 + \frac{m_{R}^2+m_{\bar{B}^{(*)}}^2-m_{B^{(*)}}^2}{2{m_{R}^2}} \left(s - m_{\bar{K}^{(*)}}^2 - m_{R}^2\right).
\end{eqnarray}  
Besides, in Ref.~\cite{Bayar:2015oea}, another set of transformation for the $s_i$ in terms of $s$ are proposed 
\begin{eqnarray}\label{Eq:MB}
   s_1 &=& \left( \frac{\sqrt{s}}{ m_{\bar{K}^{(*)}} + m_{R} }\right)^2 \left(m_{\bar{K}^{(*)}}  +  \frac{m_{B^{(*)}} m_{R}}{(m_{B^*}+m_{\bar{B}^{(*)}})}\right)^2 - \bm{p}_2^2,\nonumber\\
   s_2 &=& \left( \frac{\sqrt{s}}{m_{\bar{K}^{(*)}} + m_{R}}\right)^2\left(m_{\bar{K}^{(*)}} + \frac{m_{\bar{B}^{(*)}} m_{R}}{(m_{B^*}+m_{\bar{B}^{(*)}})}\right)^2 - \bm{p}_1^2,
\end{eqnarray}
with the consideration of the recoil.
Here the total three-momentum of the two-particle system, $\bm{p}_{1(2)}$, is estimated in terms of the binding energy of $B^{(*)}$ and $\bar{B}^{(*)}$ in the cluster $R$, 
\begin{equation}
  \bm{p}^2_{2(1)}  \approx 2m_{\bar{B}^*(B^*)}B_{2(1)}  = \frac{2 m^2_{\bar{B}^*(B^*)} m_{R}}{(m_{B^*}+ m_{\bar{B}^*})} \frac{(m_{R}+m_{\bar{K}} - \sqrt{s})}{(m_{R}+m_{\bar{K}})}.
\end{equation}
In the following, we will take this choice to evaluate the uncertainties of our prediction.

 The propagator of $\bar{K}^{(*)}$ inside the cluster, $G_0$ in Eq.~(\ref{Eq:FCAamp}), can be expressed as 
\begin{equation}
  G_0 = \frac{1}{2m_R} \int \frac{d^3 \bm{q}}{(2\pi)^3}  \frac{F_{R}(\bm{q}^2)}{{q^0}^2-\bm{q}^2 - m_{\bar{K}^{(*)}}^2 + i\epsilon}\, ,
\end{equation}
where $q_0$ denotes the energy carried by $\bar{K}^{(*)}$ in the cluster rest frame, 
\begin{equation}
  q^0 = \frac{s+m_{\bar{K}^{(*)}}^2-m_{R}^2}{2\sqrt{s}}\, ,
\end{equation}
and $F_{R}(\bm{q}^2)$ is the form factor of $B^{(*)}\bar{B}^{(*)}$ cluster, which is introduced to consider the molecular dynamics by using the Fourier transformation of the $s$-wave cluster $R$~\cite{Roca:2010tf}
\begin{eqnarray}
  F_{R}(\bm{q}^2) &=& \frac{1}{\mathcal{N}} \int_{\{|\bm{p}|, |\bm{p}-\bm{q}|<\Lambda_R\}} d^3 \bm{p} ~
  \frac{1}{4\,\omega_{B^{(*)}}(\bm{p})\omega_{\bar{B}^{(*)}}(\bm{p})}
  \frac{1}{m_{R}-\omega_{B^{(*)}}(\bm{p})-\omega_{\bar{B}^{(*)}}(\bm{p})} \nonumber\\
  && \times \frac{1}{4\,\omega_{B^{(*)}}(\bm{p}-\bm{q})\omega_{\bar{B}^{(*)}}(\bm{p}-\bm{q})} 
  \frac{1}{m_{R}-\omega_{B^{(*)}}(\bm{p}-\bm{q})-\omega_{\bar{B}^{(*)}}(\bm{p}-\bm{q})}\, ,
\end{eqnarray}
with the normalization factor $\mathcal{N}=F_R(0)$ and 
$\omega_{B^{(*)}}(\bm{p})=\sqrt{m_{B^{(*)}}^2+\bm{p}^2}$, $\omega_{\bar{B}^{(*)}}(\bm{p})=\sqrt{m_{\bar{B}^{(*)}}^2+\bm{p}^2}$.  
It is worth noting that the form factor $F_R(\bm{q}^2)$ has implicitly taken into account the interaction of $B^{(*)}$ and $\bar{B}^{(*)}$ which leads to the binding of $B^{(*)}\bar{B}^{(*)}$ system. Hence, the upper integration limit $\Lambda_R$ should take the same value of the momentum cutoff as the one used to regularize the $B^{(*)}\bar{B}^{(*)}$ loop  to get the bound state $R$.

\section{Results and discussion}\label{sec3}

\begin{figure}[b]
\begin{center}
\includegraphics[width=0.9\textwidth]{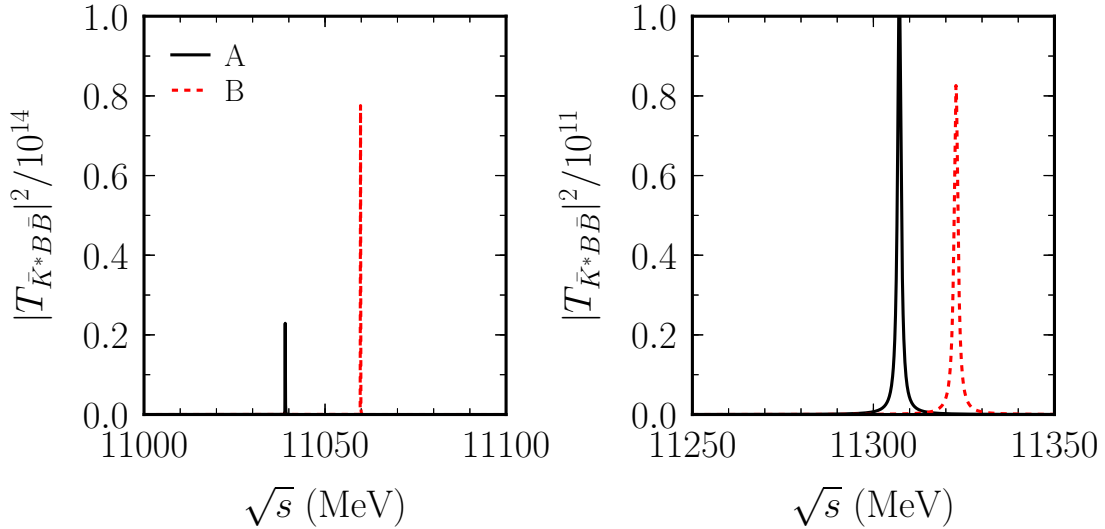}
\caption{Modulus squared of the total amplitude of $\bar{K}^* B\bar{B}$ system in $I=1/2$ with the momentum cutoff $\Lambda_R=415$ MeV. The solid (dashed) lines denote the bound states with Method A (Method B).}
\label{Fig:KxbarRamp}
\end{center}
\end{figure}

In our numerical evaluation, the meson masses are taken from Ref.~\cite{Tanabashi:2018oca} with $m_{K^*}=894.3$ MeV, $m_K=495.6$ MeV, $m_B=5279.3$ MeV, and $m_{B^*}=5324.7$ MeV.   
As mentioned in Sec.~\ref{sec2},  the form factor of cluster $F_{R}$ is regularized by a momentum cutoff $\Lambda_R$, which takes the same value as the one used in the regularization of the $B^{(*)}\bar{B}^{(*)}$  loops~\cite{Ozpineci:2013qza}. 
Generally speaking, the cutoff is a free and important parameter in the analysis of two-body bound states. 
One has to rely on some experimental information to determine/constrain it.  In Ref.~\cite{Ozpineci:2013qza}, the value of the cutoff is chosen as $\Lambda_R=415$ MeV, which is fixed to produce a bound state $X(3700)$ of the $D\bar{D}$ system~\cite{Gamermann:2006nm}. This reasonable choice for the $B$-meson sector is made to take into account that the cutoff is independent of the heavy flavor up to the order $\mathcal{O}(1/m_Q)$~\cite{Nieves:2011vw} with the heavy-quark mass $m_Q$. To estimate the errors of predicted bound states of $B\bar{B}$ and $B^*\bar{B}^*$, a range of cutoff between $415$ and $830$ MeV is used in~Ref.~\cite{Ozpineci:2013qza}. The obtained bound states of  $B\bar{B}$ and $B^*\bar{B}^*$ are rather stable while the corresponding masses are changing around 100 MeV with the cluster masses $M_{B\bar{B}}=10526$ MeV, $M_{B^*\bar{B}^*}=10616$ MeV for $\Lambda_R=415$ MeV and $M_{B\bar{B}}=10410$ MeV, $M_{B^*\bar{B}^*}=10500$ MeV for $\Lambda_R=830$ MeV. The two-body amplitudes of $\bar{K}^*B$, $\bar{K}^*\bar{B}$, $\bar{K}^*B^*$, $\bar{K}^*\bar{B}^*$ in the FCA equations are evaluated with the momentum cutoff $k_\mathrm{max} = 1070$ MeV~\footnote{We have employed the values of $k_\mathrm{max} = 1055,\, 1085$ MeV, it does not affect the results.}~\cite{Sun:2018zqs}. Since the interactions of $\bar{K}^* B$ and $\bar{K}^* B^*$ are well constrained by the observed bound states, $B_{s1}(5830)$ and $B_{s2}^*(5840)$, therefore, in the following, we will first study the $\bar{K}^*B^{(*)}\bar{B}^{(*)}$ systems and then briefly mention the results of $\bar{K}B^{(*)}\bar{B}^{(*)}$ systems.






Using the momentum cutoff $\Lambda_R=415$ MeV, in Fig.~\ref{Fig:KxbarRamp}, we present the shape of total amplitude of $\bar{K}^*B\bar{B}$ system as a function of the three-body total energy $\sqrt{s}$. To further analyze the uncertainties, as in Refs.~\cite{Debastiani:2017vhv,Bayar:2015zba}, the two schemes are used to share the three-body total energy into the two subsystems. We denote the relationship between $s_{1,\,2}$ and $s$ given in Eq.~(\ref{Eq:MA}) as method A, and another given in Eq.~(\ref{Eq:MB}) as method B.  One finds two sharp peaks at $\sqrt{s}=11039$ MeV ($11060$ MeV) and $11307$ MeV ($11323$ MeV) using method A (method B), respectively. 
Both of them are below the $\bar{K}^*[B\bar{B}]$ threshold $11420$ MeV and can be considered as the bound states of three-body system with hidden bottom. Such phenomena of two peaks was also hinted in Ref.~\cite{Dias:2018iuy}, where a bound state and a broad resonance were predicted. Besides, the difference between method A and method B is about $20$ MeV, which is consistent with the findings of Ref.~\cite{Bayar:2015zba} in the $\rho B^*\bar{B}^*$ system and Ref.~\cite{Debastiani:2017vhv} in the $DK\bar{K}$ system.

\begin{figure}[t]
\centering
\includegraphics[width=1.0\textwidth]{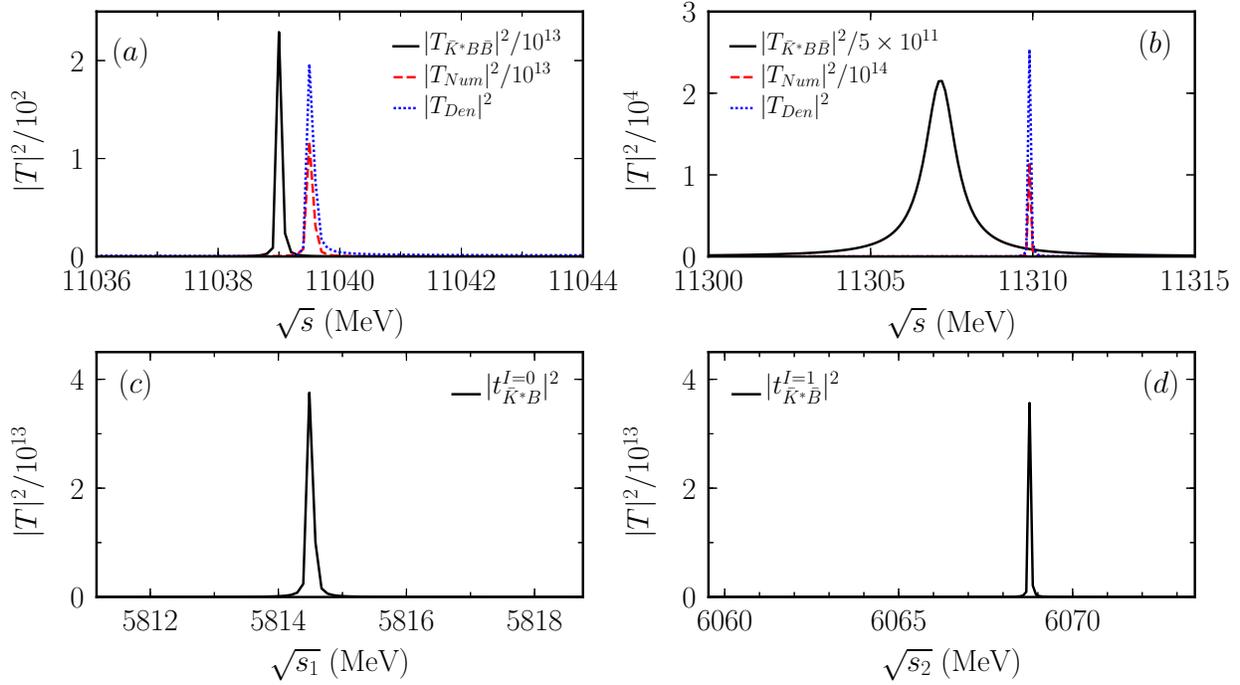}
\caption{Upper panel: The total energy dependence of the modular square of total amplitude of $\bar{K}^*B\bar{B}$ (solid lines) and the corresponding numerator (red dashed lines) and denominator (blue dotted lines) of Eq.(13). Lower panel (c): The modulus square of $\bar{K}^*B$ amplitude with isospin $I=0$ as a function of two-body c.m. energy $\sqrt{s_1}$.  Lower panel (d): The modulus square of $\bar{K}^*\bar{B}$ amplitude with isospin $I=1$ as a function of two-body c.m. energy $\sqrt{s_2}$. }
\label{Fig:TND}
\end{figure}

In order to understand the dynamical reason for producing two peaks in the shape of total amplitudes, we show the modulus square of the amplitudes of the $\bar{K}^*B\bar{B}$ three-body system, the numerator $T_\mathrm{Num}=\tilde{t}_{\bar{K}^{(*)}B^{(*)}} + \tilde{t}_{\bar{K}^{(*)}\bar{B}^{(*)}} + 2\, \tilde{t}_{\bar{K}^{(*)}B^{(*)}}\,
   \tilde{t}_{\bar{K}^{(*)}\bar{B}^{(*)}}\, G_0 $ and the denominator $T_\mathrm{Den}=1-\tilde{t}_{\bar{K}^{(*)}B^{(*)}} \, \tilde{t}_{\bar{K}^{(*)}\bar{B}^{(*)}} \, G_0^2$ of Eq.~(\ref{Eq:FCAamp}) as a function of total energy in Fig.~\ref{Fig:TND}. One can see that there are two peaks in $|T_\mathrm{Num}|^2$ and  $|T_\mathrm{Den}|^2$, which are produced from the bound state of two-body amplitudes $t_{\bar{K}B}^{I=0}$ and $t_{\bar{K}^*\bar{B}}^{I=1}$ and locates at the same $\sqrt{s}$ as the pole position of two-body peaks when transforming the  $\sqrt{s_{1,\,2}}$ to the total energy $\sqrt{s}$ using the method A, 
   as  shown in the lower panel of Fig.~\ref{Fig:TND}. 
   The total amplitude of $\bar{K}^*B\bar{B}$ is obtained 
   \begin{equation}\label{Eq:Tcom}
      T_{\bar{K}^*B\bar{B}} = \frac{T_\mathrm{Num}}{T_\mathrm{Den}} = \frac{1}{|T_\mathrm{Den}|^2} 
      \left[ T_\mathrm{Num}^{Re} T_\mathrm{Den}^{Re} + 
      T_\mathrm{Num}^{Im} T_\mathrm{Den}^{Im} -
      i\, \left(T_\mathrm{Num}^{Re} T_\mathrm{Den}^{Im}-T_\mathrm{Den}^{Re} T_\mathrm{Num}^{Im} \right)\right],
   \end{equation}
   where the superscripts $Re$ and $Im$ denote the real and imaginary parts of $T_\mathrm{Num}$ and $T_\mathrm{Den}$. Although the imaginary parts $T_\mathrm{Num}^{Im}$  and $T_\mathrm{Den}^{Im}$ are small, one cannot ignore the imaginary part of $T_{\bar{K}^*B\bar{B}}$ which also relates to the real parts of $T_\mathrm{Num}^{Re}$ and $T_\mathrm{Den}^{Re}$ as shown in Eq.~\eqref{Eq:Tcom}. This nonzero imaginary contribution will slightly change the pole position of the modulus square of $\bar{K}^*B\bar{B}$ system, such as the position of the lower pole varying from $11039.8$ MeV to $11039$ MeV and the position of the upper pole changing from $11310$ MeV to $11307$ MeV. 
   The deep bound state is produced by the $\bar{K}^* B$ interaction with isospin $I=0$ and the bound state with higher mass is originated from the $\bar{K}^*\bar{B}$ interaction with isospin $I=1$, as shown in the lower panel of Fig.~\ref{Fig:TND}. Thus, the physical picture is that the two poles correspond to the $\bar{K}^*$ sticking closer to $B$ or $\bar{B}$.

\begin{table}[t]
\caption{The masses and binding energies of the deep bound states in the $\bar{K}^*B\bar{B}$ and $\bar{K}^*B^*\bar{B}^*$ systems with the different cutoff $\Lambda_R$ (in the units of MeV).}
\label{Tab:tab1}
\begin{center}
\begin{tabular}{c|c|cc|cc|cc}
\hline\hline
 & \multirow{ 2}{*}{$I(J^P)$} & \multicolumn{2}{c|}{$\Lambda_R=415$ MeV}  & \multicolumn{2}{c|}{$\Lambda_R=830$ MeV} & \multicolumn{2}{c}{Average value} \\
\cline{3-8}
  & &  method A   & method B  & method A &  method B & Mass & Binding energy \\
 \hline
  $\bar{K}^* B\bar{B}$ & $\frac{1}{2}(1^-)$ & $11039$ & $11060$  &  $10917$  & $10992$ & $11002\, \pm 63 $ & $360\,\pm 34 $\\
  \hline
 $\bar{K}^* B^*\bar{B}^*$  & $\frac{1}{2}(3^-)$ & $11108$ & $11130$ &  $10999$  & $11076$ & $11078\,\pm 57$ & $374\,\pm 38$\\
\hline\hline
\end{tabular}
\end{center}
\label{default}
\end{table}%

\begin{table}[b]
\caption{The masses and binding energies of the heavier bound states in the $\bar{K}^*B\bar{B}$ and $\bar{K}^*B^*\bar{B}^*$ systems with the different cutoff $\Lambda_R$ (in the units of MeV).}
\label{Tab:tab2}
\begin{center}
\begin{tabular}{c|c|cc|cc|cc}
\hline\hline
 & \multirow{ 2}{*}{$I(J^P)$}  &   \multicolumn{2}{c|}{$\Lambda_R=415$ MeV}  & \multicolumn{2}{c|}{$\Lambda_R=830$ MeV}  & \multicolumn{2}{c}{Average value}   \\
\cline{3-8}
 &  &  method A   & method B & method A & method B & Mass & Binding energy \\
 \hline
  $\bar{K}^* B\bar{B}$  & $\frac{1}{2}(1^-)$ & $11307$ & $11323$ & $11180$ & $11244$ &  $11264\,\pm 65$  & $99\,\pm 28$ \\
  \hline 
 $\bar{K}^* B^*\bar{B}^*$  & $\frac{1}{2}(3^-)$ & $11377$ & $11393$ & $11259$ & $11325$ & $11339\,\pm 60$  & $114\,\pm 31$\\
\hline\hline
\end{tabular}
\end{center}
\label{default}
\end{table}%

For the three-body system $\bar{K}^*B^*\bar{B}^*$ with spins aligned ($J=3$), a similar shape of three-body amplitude with the different cutoffs is obtained.  
The results for position of peaks are summarized in Table~\ref{Tab:tab1} for the deep bound state with two values of the cutoff, $415$ MeV and $830$ MeV. One can see that the predicted bound state is quite robust although the values of the pole position are  varying by about $100$ MeV.~\footnote{This variation of the mass is an acceptable situation as shown in the $\rho B^*\bar{B}^*$ system~\cite{Bayar:2015zba}.} Such variation is major from the cutoff dependence of the cluster ($B^{(*)}\bar{B}^{(*)}$) mass as shown in Ref.~\cite{Ozpineci:2013qza}. 
The mass difference between method A and method B  is around $20$ MeV  with $\Lambda_R=415$ MeV and increases up to $\sim 80$ MeV with  $\Lambda_R=830$ MeV.
After averaging over the different results, we obtain that the mass of the deep bound state is $11002\,\pm 63$ MeV for the $\bar{K}^*B\bar{B}$ system and $11078\,\pm 57$ MeV for the $\bar{K}^*B^*\bar{B}^*$ system. In the last column of Table~\ref{Tab:tab1}, the average binding energies are also presented. Note that the binding energies of these two systems are almost the same
with $E_B=360\pm 34$ MeV and $E_B=374\pm 38$ MeV, respectively. This may look like a strong binding, but one must keep in mind that the $\bar{K}^* B$ and $\bar{K}^*B^*$ interactions are very  attractive, in fact, the produced $B_{s1}(5830)$ and $B_{s2}^*(5840)$ have the binding energies of $344$ MeV and $350$ MeV~\cite{Sun:2018zqs}. Considering the $B\bar{B}$ ($B^*\bar{B}^*$) cluster only has $33$ ($33$) MeV binding energy with $\Lambda=415$ MeV~\cite{Ozpineci:2013qza}, 
  we can conclude that the $\bar{K}^* B^{(*)}\bar{B}^{(*)}$ three-body  system is more bound than either pair, as a consequence of the combination of two subsystems. It is worth noting that a similar superbound state was also predicted in the bottom sector, $\rho B^*\bar{B}^*$ system~\cite{Bayar:2015zba}, using the same theoretical framework.
  

On the other hand, in Table~\ref{Tab:tab2}, the masses and binding energies of the heavier bound states, e.g. shown in Fig.~\ref{Fig:KxbarRamp}, are summarized for the $\bar{K}^*B\bar{B}$ and $\bar{K}^*B^*\bar{B}^*$ systems with the momentum cutoff $\Lambda_R=415$ and $830$ MeV, respectively. A similar mass difference between method A and B as Table~\ref{Tab:tab1} is observed. After performing the average over the masses and binding energies, the pole positions of the heavier states in the $\bar{K}^*B\bar{B}$ and $\bar{K}^* B^*\bar{B}^*$ systems are around $11264$ MeV and $11339$ MeV, respectively, with the binding energy around $100$ MeV. 
 


\begin{table}[b]
\caption{The masses and binding energies of the bound states in the $\bar{K}B\bar{B}$ and $\bar{K}B^*\bar{B}^*$ systems with the different cutoff $\Lambda_R$ (in the units of MeV).}\
\label{Tab:tab3}
\begin{center}
\begin{tabular}{c|c|cc|cc|cc}
\hline\hline
  & \multirow{ 2}{*}{$I(J^P)$} & \multicolumn{2}{c|}{$\Lambda_R=415$ MeV}  & \multicolumn{2}{c|}{$\Lambda_R=830$ MeV} & \multicolumn{2}{c}{Average value} \\
\cline{3-8}
  & &  method A   & method B  & method A &  method B & Mass & Binding energy \\
 \hline
  $\bar{K} B\bar{B}$ & $\frac{1}{2}(0^-)$ & $10703$ & $10722$ &  $10568$  & $10643$ & $10659\, \pm 69 $ & $305\,\pm 32 $\\
  \hline
 $\bar{K} B^*\bar{B}^*$ & $\frac{1}{2}(2^-)$  & $10953$ & $10970$ &  $10831$  & $10902$ & $10914\,\pm 62$ & $140\,\pm 32$\\
\hline\hline
\end{tabular}
\end{center}
\label{default}
\end{table}%

Furthermore, we also study the $\bar{K}B^{(*)}\bar{B}^{(*)}$ three-body systems using the FCA of the Faddeev equations. The total amplitude of the three-body interaction is determined via Eq.~(\ref{Eq:FCAamp}). The corresponding inputs of the two-body amplitudes, $t_{\bar{K}B^{(*)}}^{I=0,\,1}$, $t_{\bar{K}\bar{B}^{(*)}}^{I=0,\,1}$, are evaluated using the  effective Lagrangians Eq.~(\ref{Eq:lag}) with the local hidden gauge approach. As given in Ref.~\cite{Sun:2018zqs}, the two deep bound states, the $0(0^+)$ state with mass around 5460 MeV of $\bar{K}B$ system and the $0(1^+)$ state with mass around 5665 MeV of $\bar{K}B^*$, are predicted. These very attractive interactions will guarantee that bound states in the three-body systems are produced. In Table \ref{Tab:tab3}, we have tabulated our findings in the $\bar{K}B^{(*)}\bar{B}^{(*)}$ systems with two cutoffs, as employed in the study of  $\bar{K}^*B^{(*)}\bar{B}^{(*)}$ systems. After averaging, one finds two bound states with mass $10659\pm 69$ MeV  for the $\bar{K}B\bar{B}$ system and with mass $10914\pm 62$ MeV for the $\bar{K}B^*\bar{B}^*$ system. We also found that the difference of binding energy between $\bar{K}B\bar{B}$ and $\bar{K}B^*\bar{B}^*$ systems is around $150$ MeV, which originates from the  difference ($\sim 140$ MeV) of the binding energies of the two-body subsystems $\bar{K} B$ and $\bar{K}B^*$. 

Finally, we want to mention that, although the $\bar{K}B\bar{B}$ and $\bar{K}B^*\bar{B}^*$ systems satisfy all the criteria for a reliable application of the FCA and produce two bound states, one should take a critical look at the masses and the binding energies obtained in Table~\ref{Tab:tab3}. 
Because the two-body amplitudes of $\bar{K}B$ and $\bar{K}B^*$ determined in Ref.~\cite{Sun:2018zqs} are different with the ones from Refs.~\cite{Guo:2006fu,Guo:2006rp,Faessler:2008vc}, where the $B_{s0}^*(5725)$ and $B_{s1}(5778)$ mesons were predicted, respectively. More efforts are needed to obtain the final conclusion of the $\bar{K}B$ and $\bar{K}B^*$ interactions in order to well determine the masses of the $KB^{(*)}\bar{B}^{(*)}$ bound states.


\section{Summary}\label{sec4}

We have performed a three-body study of the $\bar{K}^{(*)}B^{(*)}\bar{B}^{(*)}$  systems using the Faddeev equations in the fixed-center approximation. The two-body subsystems $B\bar{B}$ ($B^*\bar{B}^*$) are bound forming the clusters, which interact with a light $\bar{K}^{(*)}$ meson. With the help of the observed $B_{s1}(5830)$ and $B_{s2}^*(5840)$ states, the two-body amplitudes of $\bar{K}^{*}$$B^{(*)}$ and $\bar{K}^{*}$$\bar{B}^{(*)}$ systems, used as input of the FCA equation, are well constrained in the chiral unitary approach. As a result, we found a deep bound state with mass $11002\pm 63$ MeV and a state with the higher mass $11264\pm 65$ MeV in the $\bar{K}^*B\bar{B}$ system, containing the hidden-bottom component. The similar results of two bound states with masses $11078\pm 57$ MeV and $11339\pm 60$ MeV were predicted in the $\bar{K}^* B^*\bar{B}^*$ system with spin aligned to $J=3$. Furthermore, using the constrained $\bar{K}B^{(*)}$ and $\bar{K}\bar{B}^{(*)}$ interactions by the local hidden gauge symmetry, the two bound states with $I=1/2$ are predicted in the $\bar{K}B^{(*)}\bar{B}^{(*)}$ three-body systems. We expect that the current study and Ref.~\cite{Ren:2018pcd} will arouse interest to the study of the hadronic states with hidden charm/bottom in the strange sector. 

\begin{acknowledgements}
We are grateful to Prof. E. Oset for a careful reading of the manuscript and for many valuable suggestions and comments. X.L.R. thanks the useful discussion with Prof. A. Mart\'inez Torres and Prof. K. P. Khemchandani. 
This work was partly supported by the Deutsche Forschungsgemeinschaft (DFG) and the China  National Natural Science Foundation (NSFC) through funds provided to the Sino-German CRC 110 ``Symmetries and the Emergence of Structure in QCD'' (Grant No. TRR110), NSFC (Grants No. 11775099, No. 11705069), and the Fundamental Research Funds for the Central Universities.
\end{acknowledgements}

\section*{APPENDIX}
\subsection{Two-body potentials}

The projected $J=0$ amplitudes of $\bar{K} B$ and $\bar{K} \bar{B}$ interactions: 

\begin{itemize}
\item $\bar{K}B$ and $\eta B_s$ couple channel with $I=0$
\begin{eqnarray}
	v_{\bar{K}B\to \bar{K}B}^{I=0} &=& -\frac{1}{6f_\pi^2}(2u-t-s) - 
	\frac{g^2}{2} \left(\frac{3}{m_\rho^2}+\frac{1}{m_\omega^2}\right) (s-u),\nonumber\\
	v_{\bar{K}B\to \eta B_s}^{I=0} &=& -\frac{\sqrt{6}}{12f_\pi^2}(s-u) + 
	\frac{2\sqrt{6}g^2}{3} \frac{1}{m_{K^*}^2}(s-u),\nonumber\\
	v_{\eta B_s\to \eta B_s}^{I=0} &=& -\frac{1}{36f_\pi^2}(s-2t+u).
\end{eqnarray}

\item $\bar{K}B$ and $\pi B_s$ couple channel with $I=1$
\begin{eqnarray}
	v_{\bar{K}B\to \bar{K}B}^{I=1} &=& \frac{g^2}{2} \left(\frac{1}{m_\rho^2}-\frac{1}{m_\omega^2}\right) (s-u),\nonumber\\
	v_{\bar{K}B\to \pi B_s}^{I=1} &=& -\frac{1}{12f_\pi^2}(2s-t-u) + 
	\frac{g^2}{\sqrt{2}} \frac{1}{m_{K^*}^2}(s-u),\nonumber\\
	v_{\eta B_s\to \pi B_s}^{I=1} &=& 0.
\end{eqnarray}

\item $\bar{K}\bar{B}$ single channel with $I=0$
\begin{equation}
v_{\bar{K}\bar{B}\to \bar{K}\bar{B}}^{I=0} = 
\frac{g^2}{2}\left(-\frac{3}{m_\rho^2}+\frac{1}{m_\omega^2}\right) (s-u).	
\end{equation}

\item $\bar{K}\bar{B}$ single channel with $I=1$
\begin{equation}
v_{\bar{K}\bar{B}\to \bar{K}\bar{B}}^{I=1} = 
\frac{g^2}{2}\left(\frac{1}{m_\rho^2}+\frac{1}{m_\omega^2}\right) (s-u).	
\end{equation}

\end{itemize}

The projected $J=1$ amplitudes of $\bar{K}^* B$ and $\bar{K}^* \bar{B}$ interactions:

\begin{itemize}

\item $\bar{K}^*B$ and $\omega B_s$ couple channel with $I=0$
\begin{eqnarray}
v_{\bar{K}^*B\to \bar{K}^* B}^{I=0} &=& -\frac{g^2}{2}\left(\frac{3}{m_\rho^2}+\frac{1}{m_\omega^2}\right)(s-u),\nonumber\\
v_{\bar{K}^*B \to \omega B_s}^{I=0} &=& g^2\frac{1}{m_{K^*}^2}(s-u),\nonumber\\
v_{\omega B_s \to \omega B_s}^{I=0} &=& 0.
\end{eqnarray}

\item $\bar{K}^*B$ and $\rho B_s$ couple channel with $I=1$
\begin{eqnarray}
v_{\bar{K}^*B\to \bar{K}^* B}^{I=1} &=& -\frac{g^2}{2}\left(-\frac{1}{m_\rho^2}+\frac{1}{m_\omega^2}\right)(s-u),\nonumber\\
v_{\bar{K}^*B \to \rho B_s}^{I=1} &=& g^2\frac{1}{m_{K^*}^2}(s-u),\nonumber\\
v_{\omega B_s \to \omega B_s}^{I=1} &=& 0.
\end{eqnarray}

\item $\bar{K}^*\bar{B}$ single channel with $I=0$

\begin{equation}
	v_{\bar{K}^*\bar{B}\to \bar{K}^*\bar{B}}^{I=0} = \frac{g^2}{2}
	\left(\frac{2}{m_\rho^2}-\frac{1}{m_\omega^2}\right)(s-u).
\end{equation}
\item $\bar{K}^*\bar{B}$ single channel with $I=1$
\begin{equation}
	v_{\bar{K}^*\bar{B}\to \bar{K}^*\bar{B}}^{I=1} = -\frac{g^2}{2}
	\left(\frac{1}{m_\rho^2}+\frac{1}{m_\omega^2}\right)(s-u).
\end{equation}

\end{itemize}

The projected $J=1$ amplitudes of $\bar{K}B^*$ and $\bar{K} \bar{B}^*$ interactions:
\begin{itemize}
\item $\bar{K}B^*$ and $\eta B_s^*$ couple channel with $I=0$ 
\begin{eqnarray}
  v_{\bar{K}B^*\to \bar{K}B^*}^{I=0} &=& -\frac{g^2}{2}\left( \frac{3}{m_\rho^2} + \frac{1}{m_\omega^2} \right) (s-u)\,,\nonumber\\
  v_{\bar{K}B^*\to \eta B_s^*}^{I=0} &=& -\sqrt{\frac{8}{3}}\, g^2  \frac{1}{m_{K^*}^2}  (s-u).
\end{eqnarray}
\item $\bar{K}B^*$ and $\pi B_s^*$ couple channel with $I=1$
\begin{eqnarray}
 v_{\bar{K} B^* \to\bar{K} B^* }^{I=1} &=&  -\frac{g^2}{2}\left( -\frac{1}{m_\rho^2} + \frac{1}{m_\omega^2} \right) (s-u)\,,\nonumber\\
  v_{\bar{K} B^* \to \pi B_s^* }^{I=1} &=&  \frac{g^2}{2} \frac{1}{m_{K^*}^2} (s-u).
\end{eqnarray}
\item $\bar{K}\bar{B}^*$ single channel with $I=0$ and $I=1$ 
\begin{eqnarray}
 v_{\bar{K}\bar{B}^* \to \bar{K}\bar{B}^*}^{I=0}  &=& \frac{g^2}{2} \left( \frac{3}{m_\rho^2} - \frac{1}{m_\omega^2} \right) (s-u)\,,\nonumber\\
 v_{\bar{K}\bar{B}^* \to \bar{K}\bar{B}^*}^{I=1}  &=& -\frac{g^2}{2} \left( \frac{1}{m_\rho^2} + \frac{1}{m_\omega^2} \right) (s-u).
\end{eqnarray}

\end{itemize}
The projected $J=2$ amplitudes of $\bar{K}^*B^*$ and $\bar{K}^* \bar{B}^*$ interactions:
\begin{itemize}
\item $\bar{K}^*B^*$ and $\omega B_s^*$ couple channel with $I=0$ 
\begin{eqnarray}
v_{\bar{K}^* B^* \to \bar{K}^* B^*}^{I=0} &=& -2 g^2 - \frac{g^2}{2} \left( \frac{3}{m_\rho^2} + \frac{1}{m_\omega^2} \right) (s-u)\,,\nonumber\\
v_{\bar{K}^* B^* \to \omega B_s^*}^{I=0} &=&  2 g^2 +  g^2\frac{1}{m_{K^*}^2} (s-u).
\end{eqnarray}

\item $\bar{K}^*B^*$ and $\rho B_s^*$ couple channel with $I=1$
\begin{eqnarray}
v_{\bar{K}^* B^* \to \bar{K}^* B^*}^{I=1} &=& \frac{g^2}{2} \left( \frac{1}{m_\rho^2} - \frac{1}{m_\omega^2} \right) (s-u)\,,\nonumber\\
v_{\bar{K}^* B^* \to \rho B_s^*}^{I=1} &=& 2g^2 + g^2 \frac{1}{m_{K^*}^2} (s-u).
\end{eqnarray}

\item $\bar{K}^* \bar{B}^*$ single channel with $I=0$ and $I=1$ 
\begin{eqnarray}
  v_{\bar{K}^* \bar{B}^*}^{I=0} &=& - 2 g^2 + \frac{g^2}{2}  \left( -\frac{3}{m_\rho^2} + \frac{1}{m_\omega^2} \right) (s-u)\,,\nonumber\\
  v_{\bar{K}^* \bar{B}^*}^{I=1} &=&   2 g^2 + \frac{g^2}{2}  \left( \frac{1}{m_\rho^2} + \frac{1}{m_\omega^2} \right) (s-u).
\end{eqnarray}
\end{itemize}
In the above two-body potentials, we have taken into account the width of $\rho$ meson by using the convoluted loop function. 
Besides, according to Ref.~\cite{Sun:2018zqs}, we also neglect the momentum products in the Mandelstam variable $u$, 
which corresponding to the $p$-wave contributions. 

\bibliographystyle{apsrev4-1}
\bibliography{refs}

\end{document}